\documentclass[utf8]{frontiersSCNS} % for Science, Engineering and Humanities and Social Sciences articles
\usepackage{url,hyperref,lineno,microtype,subcaption}
\usepackage[onehalfspacing]{setspace}

\def\keyFont{\fontsize{8}{11}\helveticabold }
\def\firstAuthorLast{Adhikari {et~al.}} %use et al only if is more than 1 author
\def\Authors{T. P. Adhikari\,$^{1,*}$, A. R\'o\.za\'nska\,$^{1}$, K. Hryniewicz\,$^{1}$,
B. Czerny\,$^{1,2}$, G.J. Ferland\,$^{3}$} 
\def\Address{$^{1}$Nicolaus Copernicus Astronomical Center, Polish Academy of Sciences, Bartycka~18, 00-716, Warsaw, Poland \\
$^{2}$Center for Theoretical Physics, Polish Academy of Sciences, Aleja Lotnikow 32/46, Warsaw, Poland \\
$^{3}$Department of Physics and Astronomy, The University of Kentucky, Lexington, KY 40506, USA}
\def\corrAuthor{T. P. Adhikari}

\def\corrEmail{tek@camk.edu.pl}

\begin{document}
\onecolumn
\firstpage{1}

\title[On the ILR in AGNs]{On the Intermediate Line Region in AGNs} 

\author[\firstAuthorLast ]{\Authors} %This field will be automatically populated
\address{} %This field will be automatically populated
\correspondance{} %This field will be automatically populated

\extraAuth{}

\maketitle

\begin{abstract}
In this paper we explore the intermediate line region (ILR) by 
using the photoionisation simulations of the gas clouds present 
at different radial distances from the center, corresponding 
to the locations from BLR out to NLR in four types of  AGNs.
We let for the presence of dust whenever conditions allow for 
dust existence. 
All spectral shapes are taken from the recent multi-wavelength campaigns.  
The cloud density decreases with distance as a power law.  
We found that the slope of the power law density profile  
does not affect the line emissivity radial profiles of 
major emission lines: H${\beta}$, He~II, Mg~II, C~III] ~and [O~III]. 
When the density of the cloud at the sublimation radius is 
as high as 10$^{11.5}$ cm$^{-3}$, the ILR should clearly be seen in the 
observations independently of the shape of  the illuminating radiation. 
Moreover, our result is valid for low ionization nuclear emission 
regions of active galaxies.
\section{}
\tiny
 \keyFont{ \section{Keywords:} active galaxies, emission lines, 
 photoionisation, radiative transfer, accrection disk} 
%All article types: you may provide up to 8 keywords; at least 5 are mandatory.
\end{abstract}

\section{Introduction}
The emission lines in active galactic nuclei (AGNs) provide a unique opportunity 
to study the properties of the materials located in the environment of the
supermassive black hole (SMBH). From the measurement of full width 
at half maxima (FWHM) of lines in the observed AGN spectra,  
it is well understood that there exist two separate regions of the line emission.
Lines with FWHM $\geq$ 3000 km s$^{-1}$ are emitted by materials
with densities $\sim$10$^{10}$ cm$^{-3}$ in the broad 
line region (BLR) located closer to the AGN
central engine. Whereas, the narrow line region (NLR) with gas densities 
$\sim$10$^{5}$ cm$^{-3}$, located much farther from the AGN center, 
emits the narrow lines 
with FWHM $\sim$ 500 km s$^{-1}$. A clear spatial separation in between BLR 
and NLR is present where the emission of lines with 
intermediate FWHM is not seen in the observations.

Theoretically, the lack of emission from the intermediate line region (ILR) 
was explained by  \citet[hereafter NL93]{Netzer1993}
as an effect of dust extinction, both in absorption and scattering 
of line photons and continuum. The authors considered radially distributed clouds of 
different density and ionization level. Broad and narrow line regions were separated due 
to the dust content which cannot be present in  BLR since the  gas temperatures are so 
high that the dust grains cannot survive there. Nevertheless, further out from 
the nucleus there is a boundary radius 
named sublimation radius, $R_{\rm d}$, where temperature drops substantially, 
and dust can sustain up to the distances where NLR is located.
NL93 presented that, for assumed gas parameters, the 
strong drop of line emission appears at distances where potential ILR is expected. 
Therefore natural separation between 
BLR and NLR occurs when the dust is 
taken into account in photoionisation calculations of cloud's emission. 
This natural separation disclaims the existence of the ILR. 
In the ILR, the ionisation parameter is higher than in further 
located NLR clouds and hence the relative effect of the dust absorption is stronger. 
The dust suppressed emission in ILR reappears
on transition to the NLR when dust absortion becomes negligible 
due to the low gas temperature.
However, in the recent observations of some AGNs,
additional intermediate line component of FWHM $\sim$ 700-1200 km s$^{-1}$ 
is clearly required to fit the lines in their emission spectra 
\citep{Brotherton1994,Mason1996,Crenshaw2007,Hu2008a,Hu2008b,Zhu2009,Li2015}.
The open questions are: does the ILR exist physically separated from BLR and NLR? 
What are the mechanisms that give rise to ILR in some sources but not in others?

Recently, \citet[hereafter AD16]{Adhikari2016} have shown, using the framework put forward 
by NL93, that when the density of illuminated clouds is high enough, the dust does not suppress 
the gap between BLR and NLR and intermediate line emission is clearly visible. 
The authors performed photoionisation simulations of radially distributed 
clouds subject to the radiation of four different spectral energy 
distributions (SEDs), most common types of 
AGN. The dust content was introduced at the sublimation radius of assumed value:
$R_{\rm d}=0.1$~pc. In NL93, the authors assumed a power law with 
slope -1.5 and normalization $10^{9.4}$ cm$^{-3}$ at $R_{\rm d}$ to describe the 
variation of the density of gas clouds with distance from the nucleus. These clouds 
were then illuminated by the mean AGN spectrum. 
Resulting emissivity profiles contained the suppression of 
line emission between BLR and NLR as it is commonly observed. 
AD16 made one step forward, showing that if the density at the sublimation
radius is high, of the order of $10^{11.5}$ cm$^{-3}$, the ILR is clearly visible. 
Such result appeared to be independent on SED of illuminated radiation taken into account.
Additionally, the authors argued that the low ionization nuclear emission regions (LINERs) 
should also exhibit the ILR. 

In this paper, we expand the work of AD16 and 
investigate the variation of density profile of radially distributed clouds. 
 All photoionisation simulations are done with the most recent version of 
the {\sc cloudy} code \citep{Ferland2017}.
To accommodate broad types of ionising SEDs in our 
calculations, we considered 
four distinct AGN types: Sy1.5 galaxy Mrk~509~\citep{Kaastra2011},
Sy1 galaxy NGC~5548~\citep{Mehdipour2015}, 
NLSy1 galaxy PMN~J0948+0022~\citep{Dammando2015},
and LINER NGC~1097~\citep{Nemmen2014}, each of them obtained from 
currently available simultaneous multi-wavelength observations. 
As a result of our simulations, we derived the line emissivity 
radial profiles for major emission lines:  
H${\beta}$~${\lambda}$4861.36~\AA, He~II~${\lambda}$1640.00~\AA, 
Mg~II~${\lambda}$2798.0~\AA, C~III]~${\lambda}$1909.00~\AA 
~and [O~III]~${\lambda}$5006.84~\AA. 

Adopting the density normalization to be $10^{11.5}$ cm$^{-3}$ at $R_{\rm d}=0.1$ pc, 
all the power law density distribution
yield continuous line emissivity profiles with 
prominent intermediate line emission component in 
permitted lines H$\beta$, He~II and Mg~II, independent of 
the density slopes and the spectral radiation shapes adopted. 
Below we briefly outline the photoionisation model itself, 
and discuss the resulting line emissivity profiles 
focusing mainly on the visibility of 
ILR in different AGN.

\section{Photoionisation model}
\label{sec:mod_params}
The simulation of the photoionisation process is done with the 
publicly available numerical code {\sc cloudy} version c17 \citep{Ferland2017},
which takes into account all the relevant radiative processes when a 
gas cloud is subjected to an incident radiation field. A simplistic 
geometrical set up of gas distributed form BLR further out to NLR is 
arranged by assuming spherical clouds with varying gas density, $n_{\rm H}$ ,and
the total column density $N_{\rm H}$, at each radial distances, $r$, from the SMBH: 
\begin{equation}
\label{eq:colden}
n_{\rm H} (r) = 10^{11.5} ~ (r/R_{\rm d})^{-\beta},   \,\,\,\,\,\,\,\,\,\,\,\,\   N_{\rm H}(r) = 10^{23.4} ~(r/R_{\rm d})^{-1}
\end{equation}
where $\beta$ is the power law density slope, and $R_{\rm d}$ is arbitrarily chosen
to be equal 0.1 pc (following NL93 and AD16). 
The total column density of a cloud located at  the sublimation radius  
is assumed after NL93:
$N_{\rm H}~ (\rm at~R_{d}) =10^{23.4}$ cm$^{-2}$,  and the gas hydrogen 
density after AD16 : $n_{\rm H}~ (\rm at~R_{d}) =10^{11.5}$ cm$^{-3}$. 
Here, we stress that the density normalizations lower 
than the  value adopted  in this paper, 
do not reproduce the intermediate line emission as shown by 
AD16.
 
Besides three previously considered types of AGN incident radiation shapes:
Sy1.5 galaxy Mrk~509 \citep{Kaastra2011}, Sy1 galaxy NGC 5548
\citep{Mehdipour2015}, NLSy1 PMN~J0948+0022 \citep{Dammando2015}, in this 
paper we also used the shape appropriate for LINER NGC~1097 \citep{Nemmen2014}.
This choice of SED covers the general 
shapes of the radiation emanating from the AGN central engine. 
Adopted SEDs are the incident spectra used  in the photoionisation simulation, 
where clouds distributed along the range of radii are exposed 
to the same type of radiation. All the SEDs are normalized to 
the bolometric luminosity 10$^{45}$ erg s$^{-1}$ which is an input to 
the {\sc cloudy} code. This allows to compute the 
ionizing flux i.e. ionization parameter at each cloud radius.

We adopted the {\sc cloudy} default chemical abundances, which are 
mostly the Solar values derived by \citet{Grevesse1998} for the gas
clouds at $r\leq R_{\rm d}$, whereas the interstellar medium (ISM) 
composition \footnote{for details see the Hazy1 {\sc cloudy} documentation.} 
with dust grains is used for the clouds at $ r> R_{\rm d}$. 
This assumption is consistent with the observational suggestions that 
the BLR is devoid of dust whereas the lower temperature in NLR allows 
its existence. On moving further out from BLR to NLR, the increase
in radial distance is accompanied by the decrease in ionization 
degree and a cloud thickness.  

The aim of this paper is to search how the appearance of 
intermediate line emission is sensitive on the value of 
density power law slope. Below we present line luminosity radial 
profiles for six values of $\beta$ = 0.5,1.0,1.5,2.0,2.5,3.0.
We are aware that the power law density distribution of clouds does not 
reflect realistic situation in AGN, but it is sufficient for the purpose of this paper.
In the forthcoming paper \citep[][in preparation]{Adhikari2017}, 
we plan to use realistic density profile, which is expected where clouds
form from outflowing gas above the accretion disk atmosphere. 
Furthermore, we plan to consider measured 
values of bolometric luminosities, which give the realistic position of 
sublimation radius for each type of AGN due to the formula by \citet{Nenkova2008}. 

\section{Line emissivities}
As the results of photoionisation simulations, we compute line luminosities emitted 
from clouds located at the each radii. Therefore, by presenting line luminosity 
dependence on the distance from SMBH, i.e. line luminosity radial profile, we can check 
if the emission from ILR is comparable to the BLR and NLR or substantially lower.
This is our basic test for the existence of ILR in all considered types of AGN. 

\begin{figure}[h!]
\centering
\includegraphics[width=8.5cm]{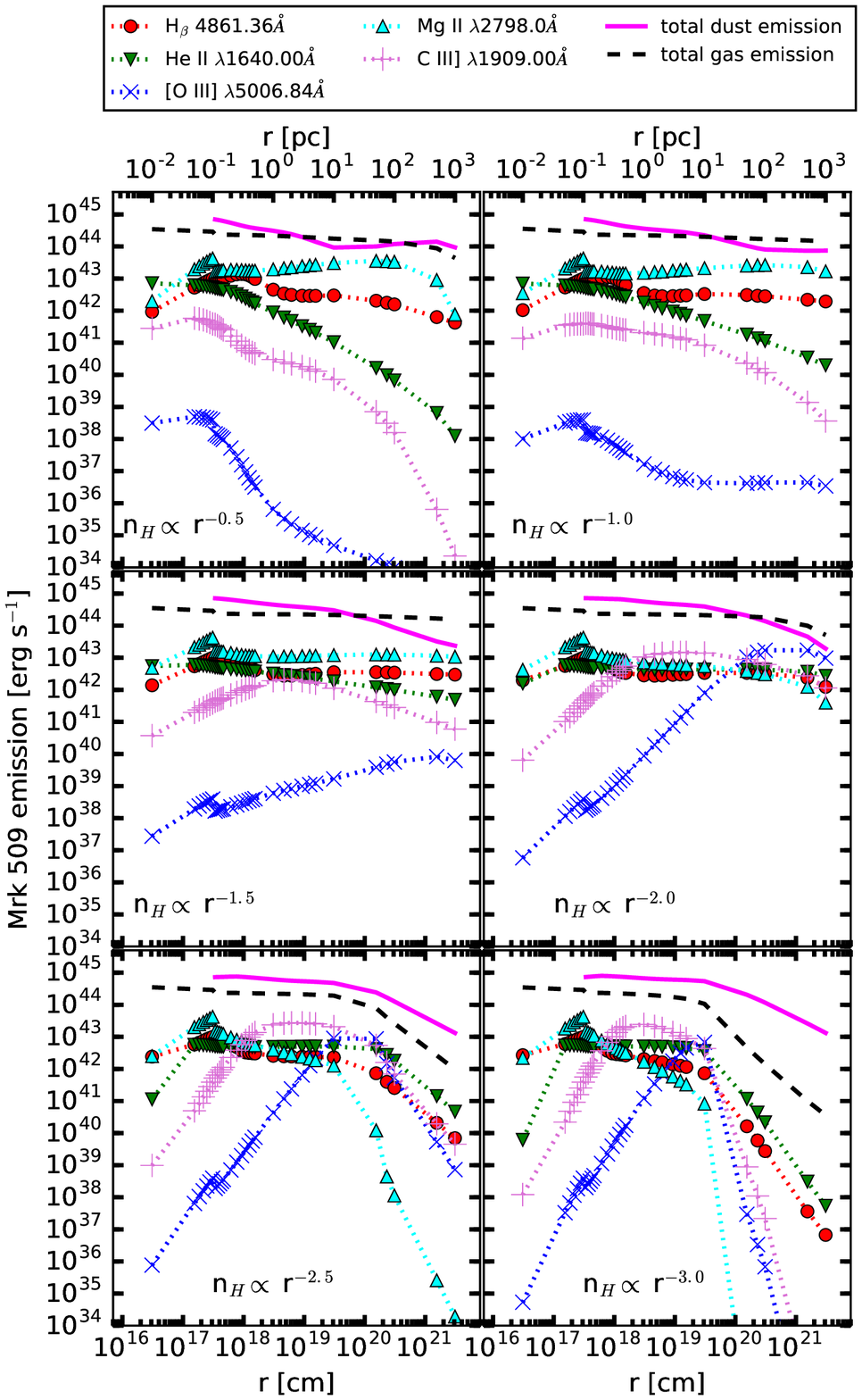}
\includegraphics[width=8.5cm]{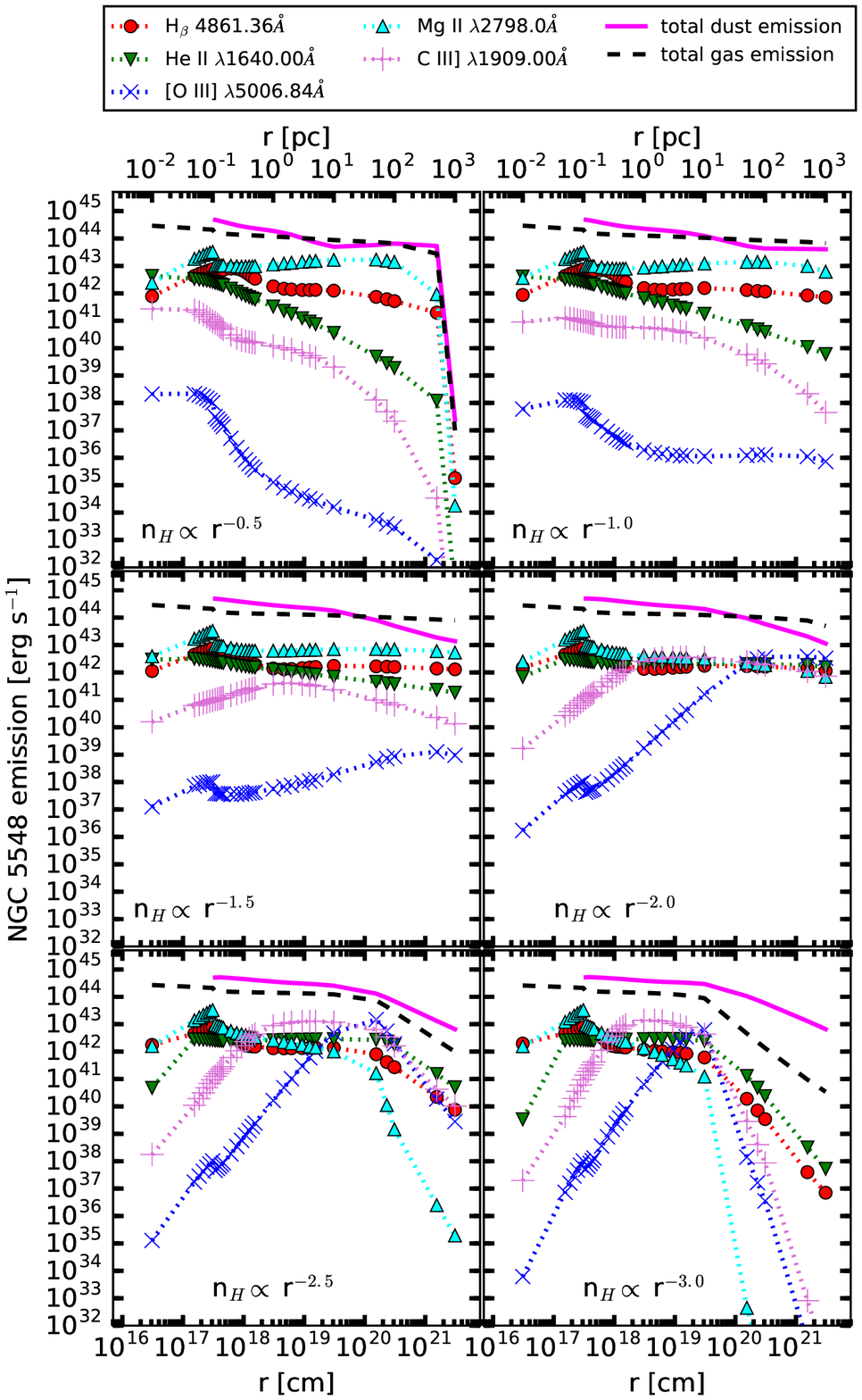}% This is a *.jpg file
\caption{\small Left panel: Line emission versus radius for 
Mrk~509 SED. Different subplots are for various density 
slopes given in the left corners.
Major emission lines: H$\beta$ (red circles), 
He~II (green triangles down), [O~III] (blue crosses),
Mg~II (cyan triangles up) and C~III] (magenta pluses) 
are shown for each density profile cases. For 
clarity total dust emission (magenta continuous line) 
and total gas emission (black dashed line) are
also shown. Right panel: The same as in left panel but with 
the spectral radiation shape of 
NGC~5548}
\label{fig:line_emission_mrk509_ngc5548}
\end{figure}

\begin{figure}[h!]
\centering
\includegraphics[width=8.5cm]{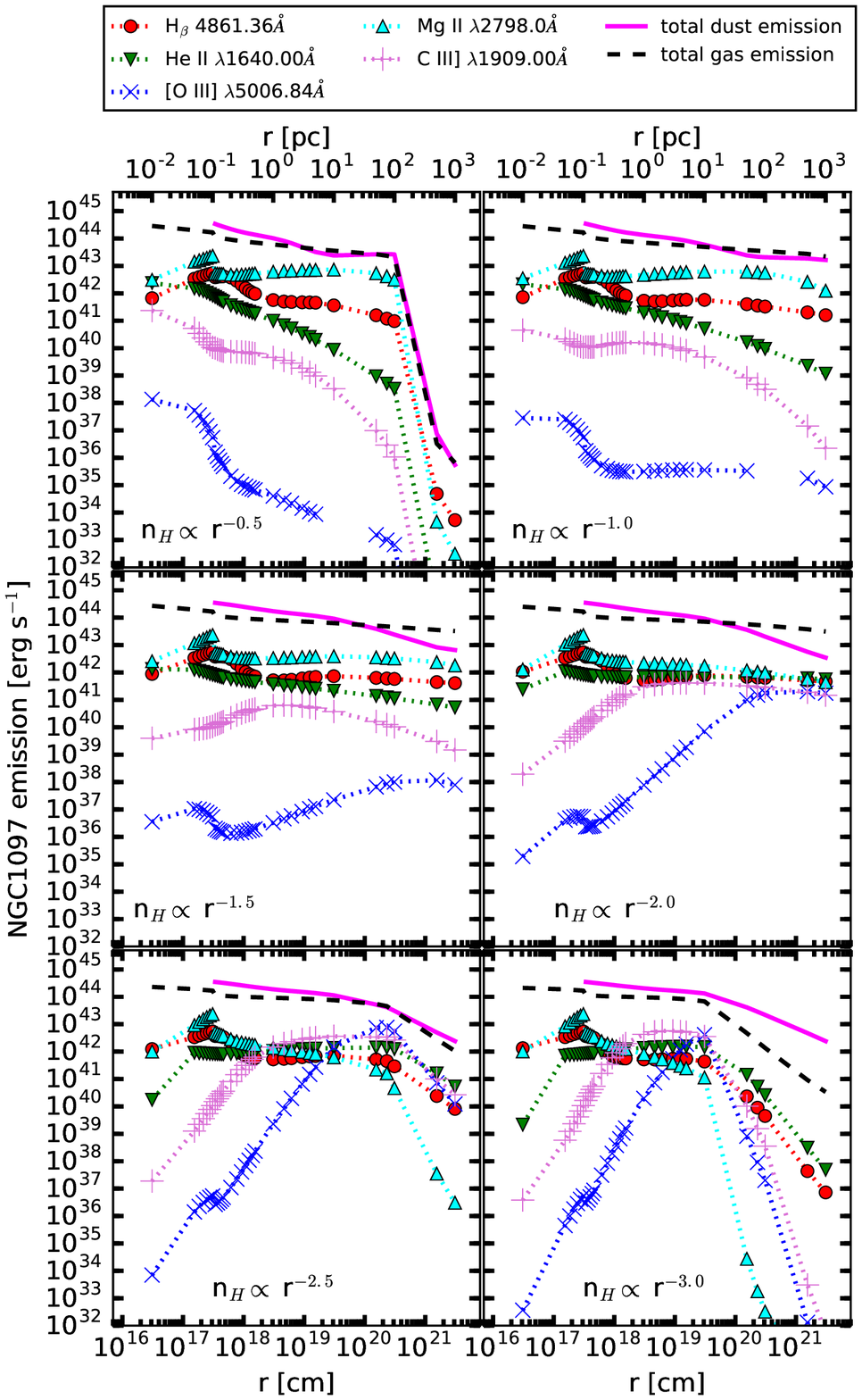}
\includegraphics[width=8.5cm]{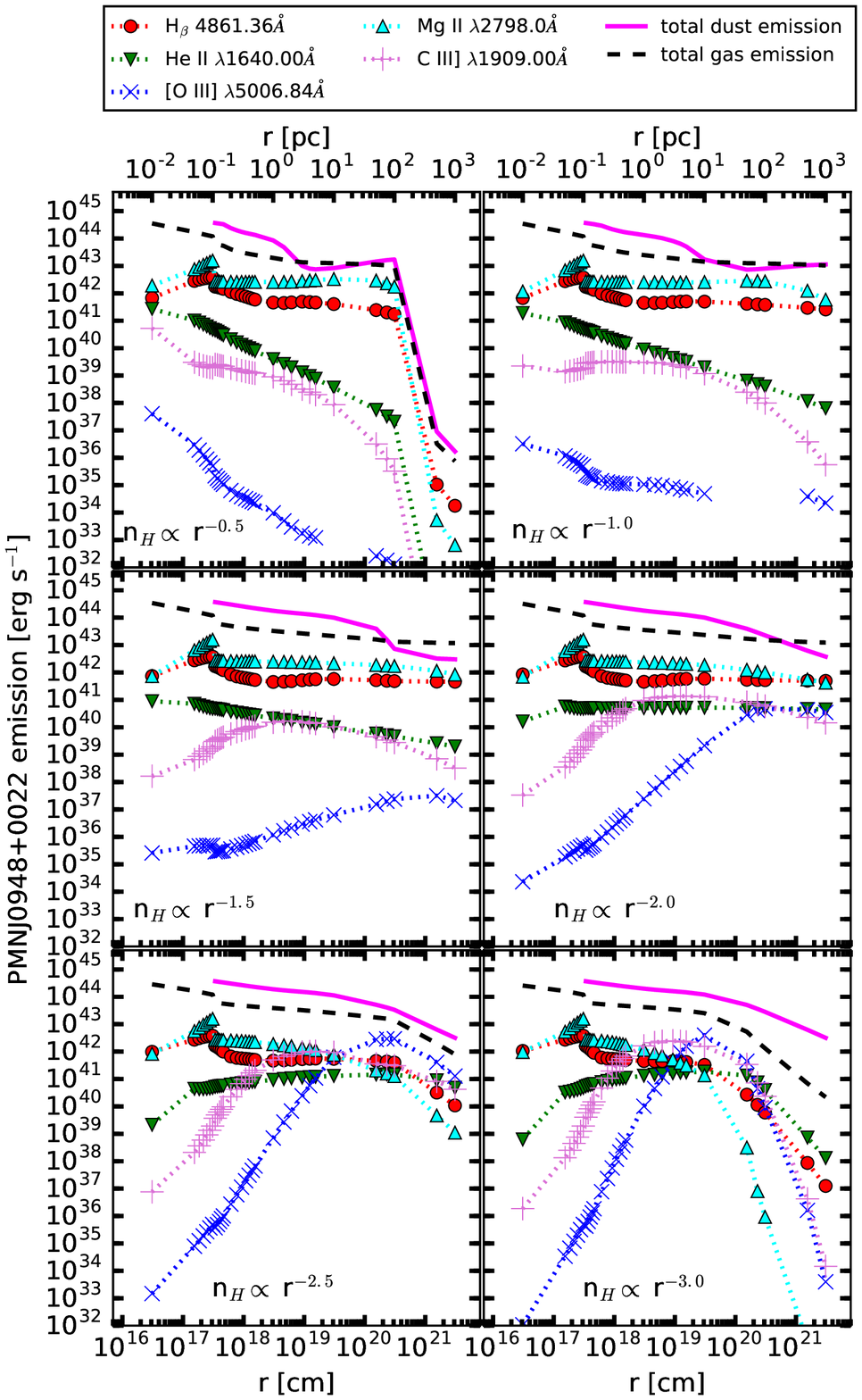}
\caption{\small The same as in Fig.~\ref{fig:line_emission_mrk509_ngc5548} but 
for the SED of NGC~1097 (left panel) and PMN~J0948+0022 (right panel).}
\label{fig:line_emission_ngc1097_PMNJ}
\end{figure}

We derived the line luminosity radial profiles  for the major emission lines
: H${\beta}$ ${\lambda}$4861.36 \AA, He~II ${\lambda}$1640.00 \AA, 
Mg~II ${\lambda}$2798.0 \AA, C~III] ${\lambda}$1909.00 \AA 
~and [O~III] ${\lambda}$5006.84 \AA. The resulting line emissivity
profiles for the four cases of SED are 
presented in the
Fig.~\ref{fig:line_emission_mrk509_ngc5548} and 
Fig.~\ref{fig:line_emission_ngc1097_PMNJ}. In all cases of density 
power law slopes, we recovered a continuous line emission, 
with a small enhancement of the
permitted lines H$\beta$ and He~II at the radial distance around 0.1 pc 
corresponding to the intermediate region, independent of the shape of
the SEDs in consideration. There is a small reduction of Mg~II line
at 0.1 pc though not very significant as compared to the 
suppression presented by NL93.
The semi forbidden line C~III] contribution to the intermediate
emission component becomes the most prominent for the density 
profile with $\beta = {\rm 1.5}$. These results corroborate 
with the conclusion of AD16 that when the density of 
the emitting gas is high enough, the extinction effect of dust 
grains on line production is negligible.

The most noticeable effects of different density profiles on 
line emissivities occur in the NLR range, i.e. for $r > 50$ pc. 
This behavior is quite obvious since for those radii, differences 
in densities between profiles are the biggest. 
For $\beta={\rm 0.5 ~and ~1.0}$, density 
falls slowly and remains moderately high across the radii
causing the strong suppression of forbidden line [O~III].
[O~III] is effectively produced in low density environment and becomes 
prominent when the density around the radius 10 pc becomes low enough,
the cases for the profiles with $\beta \geq 2.0$. 
Narrow line emission is dominated
by the C~III] components when the density distribution is given by 
the profiles with $\beta \geq 2.0$. We found that the derived line emissivities
for all cases of power law density slopes,
particularly in the region of intermediate line emission,  do
not strongly depend on the shapes of the SED used. There are subtle differences 
in emissivities corresponding to the BLR and NLR due to the different amount of UV and soft X-ray 
photons among the SEDs. This result is in agreement with the conclusion of AD16,
that the presence of  ILR emission is not determined by the shape of the 
incident radiation.

\section{Discussion}
The results above confirm the conclusion of AD16,
that the dust extinction of the emission 
lines in AGN introduced by NL93 is important 
only when the gas density is 
low. In AD16, the authors  adopted only a density power law of 
slope $\beta$ = 1.5 whereas this work 
demonstrates that the different slopes of the density distribution
do not matter significantly as long as the  gas density at the 
sublimation radius is high, in this case being 
10$^{11.5}$ cm$^{-3}$.
In all cases of considered density profiles,
we obtained an intermediate line emission around 0.1-1 pc, mostly manifested 
in permitted lines H$\beta$, He~II and Mg~II, and 
weakly present in  the semi forbidden line C~III].
This indicates that the high density and the low ionisation environment favors the 
intermediate line emission rather than the high ionization
environment where the forbidden line [O~III] is produced.
So, in the AGNs where the ILR is seen 
in observations,
the emitting region is composed of the dense and less ionized gas.
%The distance range of 0.1 -1 pc in our model, predicts the corresponding 
%reverberation mapping lags of ILR to be of the order of 100-1000
%light days, which is not yet constrained from the observations.  

The physical reason for the disappearance of the effect of dust is connected with 
the size of H- ionized front in the gas. At very high value of 
ionization parameter, i.e. for the low density case, 
the volume of the H-ionized region is very large,
if not the full cloud volume. When the density of cloud
increases, the ionization decreases, and a cloud consists of 
two regions: H-ionized region and H-neutral region. The line emission 
comes from the H-ionized region, and only the dust in this region 
competes with the gas for the photons. In other words, the dense
clouds have much smaller geometrical thickness of H -ionized layer, 
smaller dust column density in the region with abundant photons,
and therefore the dust absorption is negligible.
Our simulations are not yet aimed to make a quantitative statements 
about the studied objects. For that, we would need to do more extensive 
study, representing the bolometric luminosity and the position of the 
inner radius of the dust distribution appropriate for a given object. 
However, the grid of results shows a clear trend.

In the recent years, there has been promising claims that, broad line
emission clouds in AGN are connected with the wind from the upper part of an 
accretion disk atmosphere \citep{Gaskell2009,Czerny2011}. As shown in the 
Fig.~6 of AD16, the density profiles computed in the upper part of the standard disk 
atmosphere can be quite dense with values up to $\sim$ 10$^{15}$ cm$^{-3}$
at the assumed position of sublimation radius. Those density values depend on the
mass of the black hole and the disk accretion rate. 
Because of the high gas density, photoionisation simulations outcome with continuous 
line luminosity radial profile for the reason discussed in the previous paragraph.
As the consequence, ILR should be observed together with BLR and NLR. 
The use of realistic density profiles expected from  the 
accretion disk atmosphere is very important in the aim to understand the nature
and origin of the ILR observed in some AGNs. 
This work is in progress and will be presented
by \citet[][in preparation]{Adhikari2017}. 
\section*{Conflict of Interest Statement} 

The authors declare that the research was conducted in the absence 
of any commercial or financial relationships that could be 
construed as a potential conflict of interest.

\section*{Author Contributions}
TPA, AR and KH were responsible for developing the idea, 
doing the simulations, 
analysing the results and writing the text for the manuscript, 
BC provided the concept and GJF discussed the results. 

\section*{Funding}
This research was supported by Polish National Science 
Center grants No. 2016/21/N/ST9/03311, 2015/17/B/ST9/03422, 
2015/18/M/ST9/00541, 2015/17/B/ST9/03436, and 
by Ministry of Science and Higher Education grant W30/7.PR/2013.
It received funding from the European Union Seventh Framework Program 
(FP7/2007-2013) under the grant agreement No.312789.
TPA received funding from NCAC PAS grant for
young researchers.

\bibliographystyle{frontiersinSCNS_ENG_HUMS}
\bibliography{refs}
\end{document}